\documentclass[11pt]{llncs}
\usepackage{graphicx, color}
\usepackage{amssymb}
\usepackage{xspace}

\usepackage{times}
\usepackage{mathptm}
\usepackage{cite}

\makeatletter
\def\hb@xt@{\hbox to }
\makeatother

\let\oldendproof\endproof
\def\endproof{\qed\oldendproof}

\graphicspath{{figures/}}

\newcommand{\planarcolor}{{\sc Planar 3-co\-lo\-ra\-bi\-li\-ty}\xspace}

\newcommand{\acyclicpart}{{\sc Acyclic Partition}\xspace}
\newcommand{\naesat}{{\sc Not-all-equal-3sat}\xspace}
\newcommand{\sat}{{\sc 3-sat}\xspace}

\DeclareSymbolFontAlphabet{\Bbb}{AMSb}
\def\R{\ensuremath{\Bbb R}}

\newcommand{\red}[1]{{#1}}
\newcommand{\weg}[1]{}
\title{Self-overlapping Curves Revisited}

\author{David Eppstein\inst{1} \and Elena Mumford\inst{2}}

\institute{Computer Science Department\\
University of California, Irvine\\
\email{eppstein@ics.uci.edu}\\[0.1in]
\and Faculteit Wiskunde \& Informatica\\
Technische Universiteit Eindhoven\\
\email{e.mumford@tue.nl}}

\begin{document}

%---------------------- Text --------------------------------

\maketitle

\begin{abstract}\noindent
A surface embedded in space, in such a way that each point has a neighborhood within which the surface is a terrain, projects to an immersed surface in the plane, the boundary of which is a self-intersecting curve. Under what circumstances can we reverse these mappings algorithmically? Shor and van Wyk considered one such problem, determining whether a curve is the boundary of an immersed disk; they showed that the \emph{self-overlapping curves} defined in this way can be recognized in polynomial time. We show that several related problems are more difficult: it is NP-complete to determine
whether an immersed disk is the projection of a surface embedded in space, or whether a curve is the boundary of an immersed surface in the plane that is not constrained to be a disk. However, when a \emph{casing} is supplied with a self-intersecting curve, describing which component of the curve lies above and which below at each crossing, we may determine in time linear in the number of crossings whether the cased curve forms the projected boundary of a surface in space. As a related result, we show that an immersed surface with a single boundary curve that crosses itself $n$ times has at most $2^{n/2}$ combinatorially distinct spatial embeddings\red{, and we discuss the existence of fixed-parameter tractable algorithms for related problems}.
\end{abstract}

%\vfill\eject

\section{Introduction}
In this paper we consider the \red{algorithmic} interplay between three types of \red{topological} object: self-crossing curves in the plane, two-dimensional surfaces mapped to the plane in a self-overlapping way, and three-dimensional embeddings of surfaces that generalize the terrains familiar in computational geometry.

A \emph{surface} or \emph{two-dimensional manifold with boundary} is a compact Hausdorff topological space $M$ such that every point $p$ has a neighborhood homeomorphic to a closed disk. If this homeomorphism maps $p$ to a boundary point of the disk, we call $p$ a \emph{boundary point} of $M$; the set of boundary points is represented by $\partial M$. An \emph{immersion} or \emph{local homeomorphism} is a continuous function $i: M \rightarrow T$ that, restricted to some neighborhood of every point, is a homeomorphism. We will here be concerned only with the case that $T = \R^2$, in which case we say that the surface $M$ is \emph{immersed in the plane}. If $M$ is topologically a disk, we call $i$ an \emph{immersed disk}, but immersions of other types of manifold are also possible.

An \emph{embedding} of a surface $M$ into some space $S$ is a closed subspace of $S$ that is the image of $M$ under a one-to-one continuous function $e:M\mapsto S$, the inverse of which is also continuous.
A \emph{terrain} is a surface embedded in space $\R^3$ such that every vertical line $\{(x,y,z)\mid x=c_1, y=c_2\}$ intersects it at most once. We are interested here in a localized version of this property: an \emph{generalized terrain} is a surface $M$ embedded in space $\R^3$ such that every point of $M$ has a neighborhood the image of which is a terrain. Intuitively, such a surface is embedded in such a way that it has no vertical tangent lines, so that it has a consistent up-down orientation at every point. We will also call such a surface a \emph\emph{embedded surface}, when it does not introduce any confusion---see Figure~\ref{fi:gen-terrain} for an example. Intuitively, every generalized terrain can be constructed by gluing terrains along their boundaries. As with immersions, if $M$ is topologically a disk, we call $e(M)$ an \emph{embedded disk}.

%\begin{figure}[htb]
%\centering
%\includegraphics[width=0.4\textwidth]{xbutton} % my pdflatex doesn't like the name button.pdf
%\caption{A generalized terrain, as viewed from above.}
%\label{fi:gen-terrain}
%\end{figure}

If $i$ is an immersion, $i(\partial M)$ is a curve in the plane, which we call the \emph{boundary of the immersion}; with a suitable general-position assumption on $i$, this curve intersects itself only at proper pairwise crossings~\cite{Marx74}. And if $e$ is an embedding of a generalized terrain, we may project it into the plane to form an immersion: let $\pi_z(x,y,z)=(x,y)$, then $i(p)=\pi_z(e(p))$ is a local homeomorphism from $M$ to $\R^2$. We are interested in this paper in the conditions under which these transformations can be reversed: if we are given an immersed surface, when is it the projection of a generalized terrain? If we are given a curve in the plane, when is it the boundary of an immersed surface, or of the projection of a generalized terrain?

The study of this subject goes back to a paper by Whitney~\cite{whitney37}. Shor and Van Wyk~\cite{ShoWyk92} first considered problems of this type from the algorithmic point of view. In our terminology, they showed that it is possible in polynomial time to determine whether a given curve (with proper pairwise crossings) is the boundary of an immersed disk. However, the possibilities of non-disk manifolds, of curves with multiple connected components, and of space embeddings as well as of plane immersions left many similar types of problems unsolved.

\begin{figure}[t]
\centering
\includegraphics[width=5in]{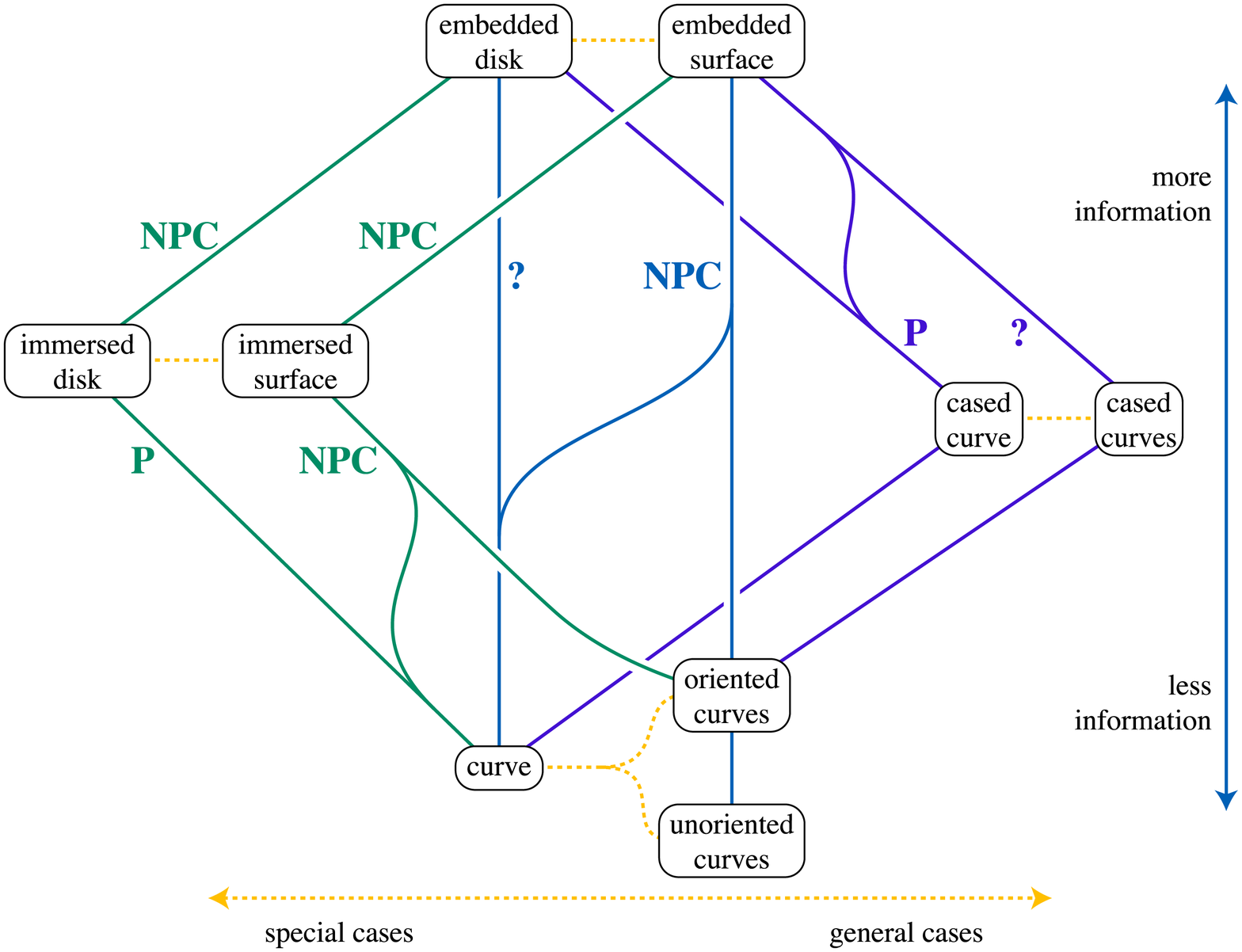}
\caption{}
\label{fi:roadmap}
\end{figure}

Figure~\ref{fi:roadmap} depicts the set of objects whose relationships we consider; a transformation from one type of object to another that loses information about the object is depicted as a downward arc. The algorithmic problems we consider, therefore, correspond to the upward arc in this figure; for instance, the arc from curves to immersed disks, labeled ``P'', represents the result of Shor and Van Wyk that one can determine in polynomial time whether a curve is the boundary of an immersed disk; they call a curve that has this property \emph{self-overlapping}. The remaining labels on the arcs of the figure represent our new results.

The cased curves mentioned on the right side of Figure~\ref{fi:roadmap} require further explanation. If one computes the projected boundary of a generalized terrain, one can obtain not just a set of curves in the plane, but also a ``casing'' describing the above-below relationship of the two components of boundary curve at each crossing point. Casings are generally depicted graphically by interrupting the lower curve segment at a crossing, and allowing the upper curve segment to pass through the crossing uninterrupted, as we have done with the arcs and crossings of the figure. Casings of this type are standard in the two-dimensional description of mathematical knots, and have also been studied from the point of view of graph drawing~\cite{EppKreMum-WADS-07}. By throwing away the casing information, we obtain an uncased curve in the plane, but an uncased curve with $n$ crossings has $2^n$ different casings. As we show, this casing information is crucial: with it we can, in linear time, reconstruct a generalized terrain (if one exists) that has the given cased curve as its projected boundary. In the absence of casing, as we show, it is NP-complete to determine whether a curve is the projected boundary of a generalized terrain, or the boundary of an immersed surface. Additionally, we show using a different reduction that it is NP-complete to determine whether an immersed disk is the projection of a generalized terrain. The question mark indicates the problem that remains open.

The problem of embedding a cased curve as a surface boundary is related to the study of so-called Seifert surfaces in the knot theory; however, we require the surface to be a generalized terrain, which is not generally the case for Seifert surfaces. Two spatial embeddings of a curve are called \emph{combinatorially equivalent} if they provide it with the same casing. Thus, every curve with $n$ crossings has at most $2^n$ combinatorially different spatial embeddings. However, we show that an immersed surface has at most $2^{n/2}$ combinatorially different spatial embeddings. \red{As we show, the ideas behind this combinatorial result lead to a fixed-parameter-tractable algorithm for testing whether an immersed surface can be lifted to an embedded surface.}

\red{Our main reasons for being interested in these problems are, firstly, their mathematical pedigree stretching back to Whitney's 1937 paper and, secondly, to resolve past confusion over how variations in the definition of a self-overlapping curve can change the computational complexity of the problem. We note for instance that the original conference version of the Shor and Van Wyk paper stated incorrectly that their methods solved both the immersion problem and the (still open) problem of testing whether a curve is the boundary of an embedded disk; this error was fortunately corrected in the journal version. Beyond these motivations, however, it also may be possible to use these problems in certain application areas of geometric computation. For instance, VLSI circuits are typically designed in terms of two-dimensional polygonal regions that overlap with each other in functional regions of the circuit; the global vertical ordering of these regions is less critical for the proper operation of the circuit than the vertical adjacencies between overlapping regions. Thus, our results may shed light on the difficulty of testing whether certain two-dimensional circuit designs have a valid three-dimensional representation.  In architectural and landscape design, also, one often has surfaces with no vertical tangent (because people must walk on them!) with levels that overlap in complex patterns. More generally, we feel that the concept of a generalized terrain may be relevant in many of the computational geometry applications in which terrains arise.}

\section{Additional definitions and examples}

Let $i: M \rightarrow \R^2$ be a surface immersion in the plane. For every point $p \in i(M)$ the \emph{thickness} of $i(M)$ at $p$ is the number of points in the set $i^{-1}(p)$.  The boundary of an immersed surface splits $\R^2$ into faces, where the thickness at all points belonging to the same face is the same. In fact, the thickness at $p$ can be obtained from the boundary curve of the surface and is equal to the \emph{winding number} of the curve around $p$, defined as the number of times the curve goes around $p$ in clockwise direction; see, e.g.~\cite{ChiSte-66} for a more detailed explanation of this concept.

\begin{figure}[t]
\begin{minipage}[t]{.45\textwidth}
\centering
\includegraphics[width=0.9\textwidth]{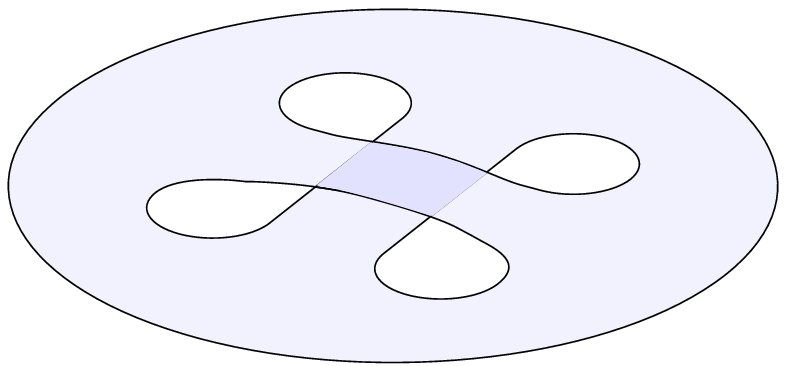} % my pdflatex doesn't like the name button.pdf
\caption{A generalized terrain, as viewed from above.}
\label{fi:gen-terrain}
\end{minipage}
\hfill
\begin{minipage}[t]{.45\textwidth}
\centering
\includegraphics[width=0.9\textwidth]{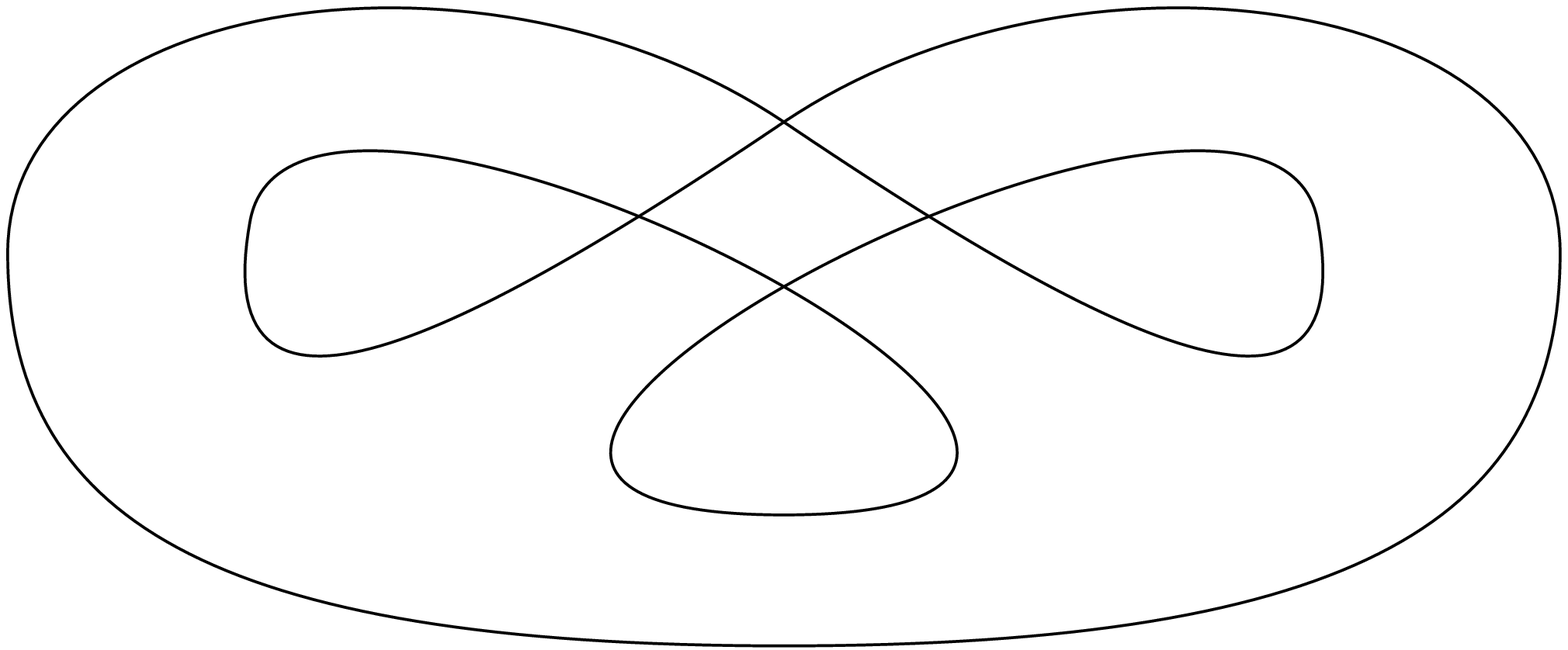}
\caption{This curve is the boundary of a unique immersed surface, topologically equivalent to a punctured torus.}
\label{fi:pretzel}
\end{minipage}
\end{figure}

A \emph{lifting} of an immersion $i$ is an embedding $e:M\rightarrow \R^3$ that projects to $i$: that is, for all $p\in M$, $i(p)=\pi_z(e(M))$, where $\pi_z((x,y,z))=(x,y)$. Necessarily, $e$ must describe a generalized terrain, for otherwise its projection would not be a local homeomorphism.

A \emph{hole} in a surface $M$ with an immersion $i$ is a component $C$ of the boundary of $M$ such that $i(C)$ is a simple curve and such that $i$ maps a neighborhood of $C$ to the outside of $i(C)$. It is tempting to imagine that every immersed surface, and therefore every generalized terrain, must be topologically equivalent to a disk with holes, and that every immersed surface with a single boundary component must be topologically equivalent to a disk, but this is not true. For instance, Figure~\ref{fi:pretzel} shows a single curve that bounds an immersed surface topologically equivalent to a punctured torus.

In dealing with surfaces that have multiple boundary components, it is important to have the concept of an \emph{orientation} of a curve, an assignment of a consistent cyclic ordering to the points of the curve. We orient the boundary components of a surface $M$ so that (as viewed according to the orientation) the surface itself is to the right of the boundary curve. An immersion or embedding of a surface, having a given set of curves in the plane as boundary, is consistent with an orientation of those curves if this rightwards orientation intrinsic to the surface matches the orientation in the embedding. Thus, a simple closed curve bounding a disk in the plane is oriented clockwise, while the boundary of a hole is oriented counterclockwise.

\begin{figure}[b]
\begin{minipage}[t]{.45\textwidth}
\centering
\includegraphics[width=0.75\textwidth]{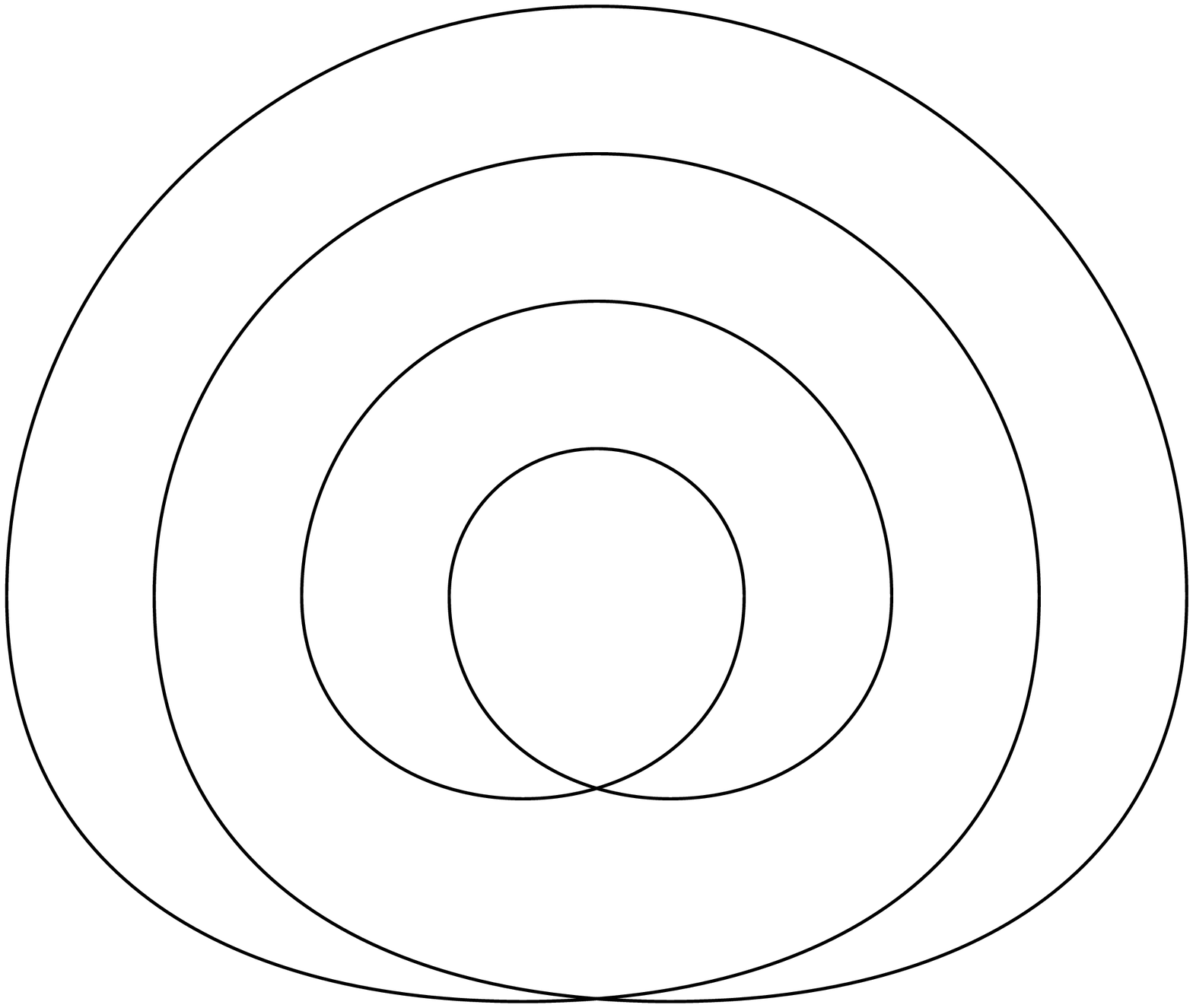}
\caption {These two curves form the boundary of an immersed surface in two different ways, neither of which can be lifted to $\R^3$.}
\label{fi:doubletwist}
\end{minipage}
\hfill
\begin{minipage}[t]{.45\textwidth}
\centering
\includegraphics[width=0.6\textwidth]{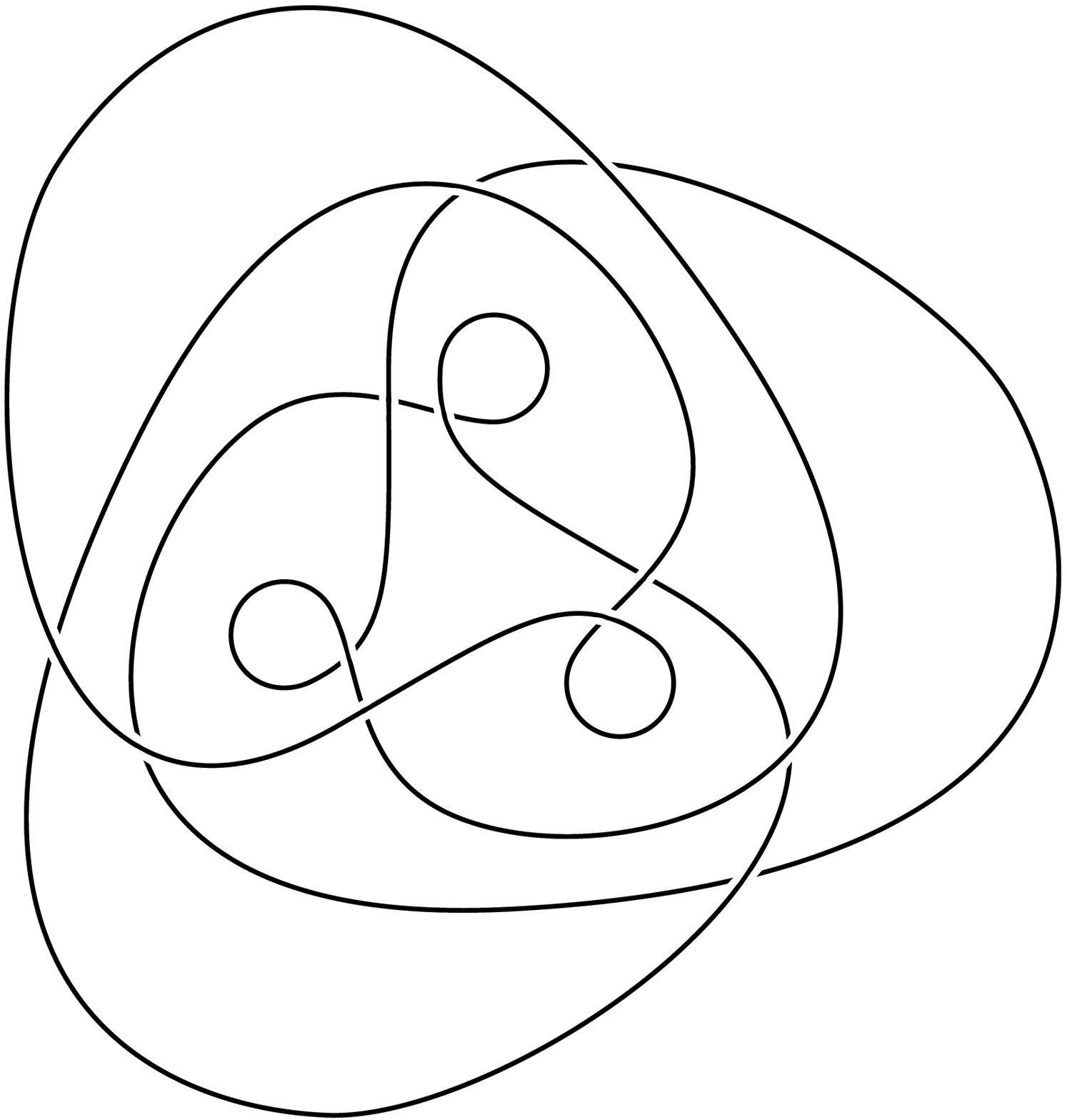}
\caption{Bennequin's curve bounding five different immersed disks.}
\label{fi:Bennequin}
\end{minipage}
\end{figure}

Although every generalized terrain projects to an immersed surface and every immersed surface has a self-intersecting curve as a boundary, this correspondence does not always work in the opposite direction. Some immersed surfaces cannot be be lifted into space; Figure~\ref{fi:doubletwist} shows an example of an immersed disk with a hole. Any embedding of this disk into $\R^3$ would intersect itself at some curve  connecting the self-intersection points of its boundary curve.  Some curves, such as the curve defined by the $\infty$~symbol, cannot be embedded as boundaries of any immersed surfaces. Even more problematically, some curves can be a boundary of more than one immersed surface. Bennequin in \cite{Bennequin75} has given an example of a curve (see Figure~\ref{fi:Bennequin}) that can be viewed as a boundary of an immersed disk in five different ways. Two of these ways involve a single central component with three lobes hanging off it symmetrically, while the other three have a shape that is more like a single spiral strip. The three spiral disks can be embedded into space but the two three-lobed disks cannot.

\section{Lifting a cased curve}

\begin{theorem}
\label{cased alg}
Given a single component curve with a given casing, we can determine whether the cased curve represents the boundary of a generalized terrain, and find a terrain having it as boundary, in time linear in the number of crossings of the curve.
\end{theorem}
\begin{proof}
We represent the curve using a modified form of Dowker notation~\cite{Dowker83} as follows:
\begin{enumerate}
\item Label the $n$ crossings of the curve by the numbers from $1$ to $n$, arbitrarily.
\item Choose a starting point on a part of the curve that lies on the exterior face of the drawing, and orient the curve consistently so that at this starting point the exterior face lies to the left of the oriented curve.
\item At crossing $i$, draw two arrows, one on each of the two strands of the curve, outward from the crossing in the direction given by the orientation. Assign the arrow that is to the clockwise of the other arrow the number $+i$ and the other the number $-i$.
\item List the marks constructed in this way, in the order that they would be found by traversing the curve as oriented from the given starting point.
\end{enumerate}
Additionally, we specify the casing as an $n$-bit binary number. It has a nonzero bit in position $i-1$ if the strand labeled $+i$ crosses above the strand labeled $-i$, and a zero bit otherwise.

We use the sequence of above-below relationships to compute an array $Height[i]$, that indicates the number of layers of surface that are supposed to lie below the curve as it heads outwards from the crossing labeled $i$. Note that here, $i$ may be either positive or negative. At the start of the labeling sequence, the height is zero: as it is on the outside face, the surface must be only one layer thick at that point. Subsequent heights can be computed in constant time per value, by traversing the curve: at an undercrossing, the height remains the same as at the previous arrow in the traversal order, while at an overcrossing, the height differs either by $+1$ or $-1$. For an overcrossing with label $+i$, the new height is one larger than the previous height, but for an overcrossing with label $-i$, the new height is one smaller than the previous height.

From these heights of the boundary curve, we can define a full surface, by stacking sheets of surface above each face of the drawing, the number of sheets equalling the winding number of the curve around the face.
The height of the boundary curve tells us which sheets continue from one face to an adjacent face, and which sheet has its boundary there. This describes a valid surface if the following conditions are met:
\begin{enumerate}
\item The height of each strand of the curve lies between zero and the winding number of the face adjacent to the strand in the exterior direction. (These winding numbers may be determined from the curve itself in linear time, based on a similar traversal of the crossing sequence.)
\item The casing determined by the heights matches the input casing.
\end{enumerate}
Both conditions may easily be checked in linear time.
\end{proof}
For a curve with multiple components we have a similar result but the complexity of the algorithm is higher.
\begin{theorem} Given a cased oriented curve, we can determine whether the cased curve represents the boundary of a generalized terrain and find a terrain having it as a boundary, in $O(min(nk,n+k^3))$ time, where $n$ is the number of self-crossings of the curve and $k$ is the number of its components.
\end{theorem}
\begin{proof}
We first calculate the winding numbers of the faces our curve $C$ splits the plane into. For that we construct a \emph{dual graph} $D(C)$ of the curve defined as follows. $D(C)$ has a vertex for each face. Two vertices $v_1$ and $v_2$ of $D(C)$ are connected by a directed edge $(v_1\rightarrow v_2)$ if the corresponding faces $f_1$ and $f_2$ are separated by a directed strand of $C$ in such a way that $f_1$ lies to the left as we move along the strand in the given direction. Let $v$ have a winding number $w$. Then every neighbor $u$ of $v$ has a winding number $w-1$ if the corresponding edge is oriented from $u$ to $v$ or has the winding number $w+1$ otherwise. (If \weg{both $(u,v)$ and $(v,u)$ edges are present, or} we encounter a vertex with a negative winding number, the curve cannot be embedded as a generalized terrain boundary.) The winding number at the vertex representing the outer face is zero and $D(C)$ is connected, so we can find winding numbers for all vertices of $D(C)$ in time linear in $n$.
%DAVID:
Next, as in the single-curve algorithm, we pick a starting point for every curve component $c_i$ with as low thickness (the winding number of a face adjacent to  the left of $c_i$) as possible; let $b_i$ denote the height of the curve $c_i$ at that starting point. By tracing around the curve, as before, we can determine the height of the curve at each of its other strands $s$, as an integer offset $\gamma_{s}$ from $b_i$. To ensure correctness of the casing, we have four types of constraints on these heights:
\begin{description}
\item{(1)} $b_i+\gamma_{s}$ values are non-negative for all strands $s$ of $c_i$, for all $0 \leq i \leq k$. This will follow automatically from our choice of start point as long as $b_i$ itself is non-negative.
\item{(2)} Each $b_i+\gamma_s \leq w$, where $w$ is the winding number adjacent to $s$ (on its left). We put all constraints of this type for a single component together to find the smallest integer $B_i$ such that $b_i$ must be at most $B_i$ in order to satisfy the given constraints.
\item{(3)} Each crossing between two strands $s_1$ and $s_2$ of the same component $c_i$ is cased correctly. This can be tested by comparing their offsets $\gamma_{s_1}$ and $\gamma_{s_2}$, and does not depend on the choice of $b_i$.
\item{(4)} Each crossing between two different components $c_i$ and $c_j$ is cased correctly. This translates to the inequality $b_i \geq b_j - \delta_{ij}$, for some value $\delta_{ij}$ that can be calculated from the two offsets $\gamma_{s^i}$ and $\gamma_{s^j}$ at the two crossing strands $s^i \in c_i$ and $s^j \in c_j$.
\end{description}
To handle constraints of types (1), (2), and (3), make a graph $G$, with one vertex $v_i$ per component $c_i$, and a special starting vertex $s$. Draw an edge with length zero from $s$ to each $v_i$, and an edge of length $\delta_{ij}$ from $v_j$ to $v_i$. If there exists a negative cycle in this graph, it corresponds to a set of constraints of type (4) that cannot be simultaneously satisfied, and no embedding exists. Otherwise, let $b_i$ be  $-d(s,v_i)$ where $d$ is the distance computed by a single source shortest path algorithm (note the minus sign). The edge from $s$ to $v_i$ forces $d(s,v_i)$ to be non-positive, so $b_i \geq 0$. The  edge from $v_j$ to $v_i$ forces $d(s,v_i) \leq d(s,v_j) + \delta_{ij}$,  the negation of a constraint of type (4), so all such  constraints are satisfied. We can test whether this assignment  of $b_i$ values satisfies the constraints of type (2) by testing each such inequality; if one of the constraints of this type is  violated then it together with the constraints on the shortest  path to $v_i$ cannot be simultaneously satisfied and no embedding  exists. Therefore, the casing corresponds to an embedding if and only if setting $b_i = -d(s,v_i)$ results in a height  assignment that satisfies all constraints.  If there are k curve components, the graph has $O(min(n,k^2))$ edges, the negative cycle detection and single source shortest path in graph with negative edge weights can be done in $O(min(nk,k^3))$ time (e.g. by Bellman-Ford~\cite{Bellman58,Ford62}) so the total time would be  $O(min(nk,n +k^3))$.
\end{proof}
This result immediately leads to an $O(n2^n)$-time algorithm, which we have implemented, for testing whether an uncased curve is the boundary of a generalized terrain: simply apply this linear time test to all possible casings of the curve. However, we believe that dynamic programming based on a separator decomposition of the curve arrangement will lead to improved running times. In such a technique, similarly to dynamic programs for other planar graph algorithms described by Smith and Wormald~\cite{SmiWor-FOCS-98}, one would partition the planar graph representing the input curve arrangement along a separator, and maintain a system of dynamic program states describing the heights of points on the input curve at each position at which the separator intersects the input curve.  As the heights are at most $n$, and each dynamic programming state would need to combine the heights of $O(\sqrt n)$ points, the time for such an approach would be exponential in $O(\sqrt n\log n)$, improving the simple algorithm described above which is exponential in $n$. However, such an algorithm would be significantly more complicated than the one we implemented.

\section{Hardness of immersion and embedding for uncased curves}

\begin{figure}[t]
\centering
\includegraphics[width=0.85\textwidth]{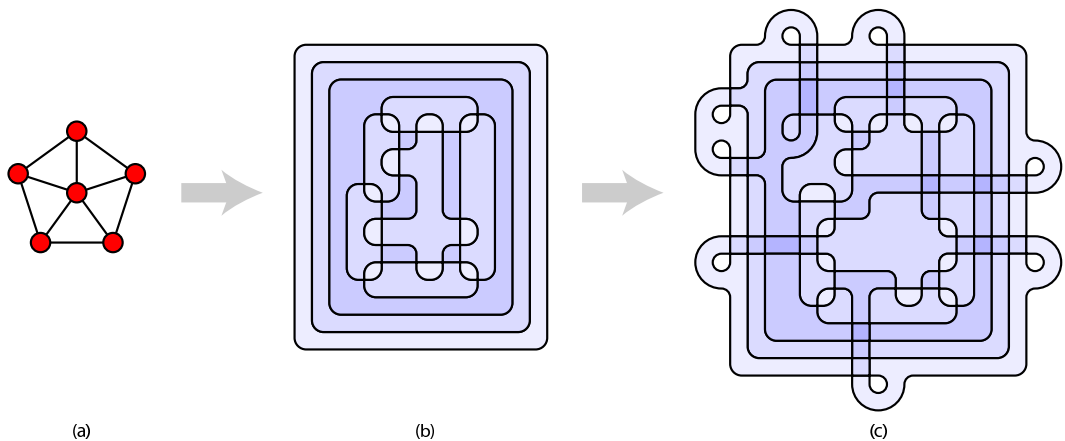}
\caption{Reduction from \planarcolor to immersibility or embeddability of a curve.}
\label{fi:embedcurve-NP}
\end{figure}

As we now show, the absence of a casing makes it much more difficult to determine whether a given curve or set of curves is the boundary of an immersion or of a generalized terrain. The overall reduction is depicted in Figure~\ref{fi:embedcurve-NP}. We start with a simplified version of the proof that applies to multiple oriented curves.

\subsection{Oriented curves}\label{se:oriented}
We show that finding a surface immersion for an oriented curve is NP-complete using a reduction from {\planarcolor}. A {\planarcolor} instance tests whether a planar graph has a vertex-coloring with three colors, such that adjacent vertices have different colors.

We transform a planar graph $G = (V,E)$ into a family $C_G$ of $|V|+3$ simple closed curves. $|V|$ of these curves, which we call vertex-curves, are oriented counterclockwise and represent the vertices of $G$. In any immersion or embedding having $C_G$ as boundary, vertex-curves must be hole boundaries due to their orientation. We place these curves in such a way that every pair of vertex-curves are either disjoint or have two points in common, and such that a pair of vertex-curves intersect if and only if the corresponding vertices of the graph are adjacent. The remaining three curves in $C_G$, denoted $c_1$, $c_2$ and $c_3$, are oriented clockwise and surround all the vertex-curves. These three curves are disk-boundaries. See Figure~\ref{fi:embedcurve-NP}(a),(b) for an example.
\begin{lemma}
\label{3c => is}
If $G$ has a three-coloring then there exists an immersed surface in the plane of which $C_G$ is a boundary, having the topology of three disks with holes.
\end{lemma}
\begin{proof}
Form a disk in the plane for each of $c_1$, $c_2$, and $c_3$. For each vertex-curve $c_v$, corresponding to a vertex $v$ with color $i$, form a hole with boundary $c_v$ in the disk bounded by $c_i$. Due to the coloring of $G$, no two holes in the same disk overlap, so this forms a valid immersion of three punctured disks with boundaries $c_1$, $c_2$ and $c_3$, together with the holes formed as above.
\end{proof}
\begin{corollary}
If $G$ has a three-coloring then $C_G$ can be embedded in space as the boundary of a generalized terrain.
\label{cor:col-embedding}
\end{corollary}
\begin{proof}
Lift the three punctured disks of Lemma~\ref{3c => is} to distinct heights in $\R^3$.
\end{proof}
Next we prove that given a collection of surfaces $S$ immersed in the plane so that the boundaries of $S$ match the reduction $C_G$ of a graph $G$ we can define a $3$-coloring of $G$ from $S$.

To do so, define a relation $\sim$ between hole-boundaries of $S$: $b_1\sim b_2$ if and only if there exists a curve in $S$ that starts on $b_1$, ends on $b_2$, and does not pass through any points where two holes overlap.
\begin{lemma}
$\sim$ is an equivalence relation.
\end{lemma}
\begin{proof} Reflexivity and symmetry are clear. To prove transitivity, suppose $b_1 \sim b_2$ and $b_2 \sim b_3$, and find a curve from $b_1$ to $b_3$ by concatenating the curves $b_1 \sim b_2$, $b_2 \sim b_3$, and a curve around the boundary of $b_2$.
\end{proof}
\begin{lemma}
$\sim$ has at most three equivalence classes.
\end{lemma}
\begin{proof}
We pick a point $p$ of the plane that is not in any of the holes: the immersed surface has three points $p_1$, $p_2$, $p_3$ that map to $p$. Then every point of the surface can be connected by a path to one of the three points $p_i$. The path is constructed by first following a straight line segment to $p$ and then detouring around any hole crossed by the line segment. Any two hole boundaries that connect to the same $p_i$ must be equivalent to each other by concatenation of curves.
\end{proof}
\begin{lemma}
If hole-boundaries $b_i$ and $b_j$ cross each other, then $b_i$ is not related by $\sim$ to $b_j$.
\end{lemma}
\begin{proof} Suppose for a contradiction that $b_i \sim b_j$, let $c$ be a curve connecting $b_i$ and $b_j$, and let $s$ be a short line segment connecting $b_i$ to $b_j$ near their crossing. We can assume that $c$ and $s$ have the same endpoints, so together they define a (possibly  self-intersecting) polygon, containing some other set of boundary holes. We choose $c$ in such a way that this polygon contains as few boundary holes as possible, and then (among curves passing around the other holes in the same pattern) so that it is as short as possible. If $c$ is not equal to $s$, then it must be stretched tight against some other hole, and we can reduce the number of holes in the polygon by replacing $c$ by a curve that goes the other way around the same hole. Thus $c$ must be equal to $s$. But then near the crossing $b_i$ and $b_j$ would be on the same layer of the surface, not possible.
\end{proof}

Thus $\sim$ corresponds to a partition of $V$ into three subsets, with no adjacent pair of vertices in the same subset; that is, a 3-coloring. $G$ may be transformed into $C_G$ in polynomial time, providing a polynomial-time many-one reduction from the known NP-complete problem of {\planarcolor} to immersing or embedding a set of oriented curves. These immersion and embedding problems may be solved in NP, in the case of embedding by considering heights of curves as used in the proof of Theorem~\ref{cased alg}, or in the case of immersions by a system of disks and gluings as used by Shor and Van Wyk~\cite{ShoWyk92}.
We have proven
\begin{theorem} The problem of determining whether a set of oriented curves can be seen as a boundary of some immersed surface is NP-complete.
\label{le:multi-lift}
\end{theorem}
\begin{corollary}
The problem of determining whether a set of oriented curves can be seen as a boundary of an embedded surface is NP-complete.
\end{corollary}
\begin{proof}
Consider a planar graph $G$ and the set of curves $C_G$ described above. Assume we have a generalized terrain $M$ that has $C_G$ as a boundary. $M$ projects to an immersed surface with $C_G$ as a boundary which, as we have shown earlier, defines a three coloring of $G$. In the other direction, by Corollary~\ref{cor:col-embedding}, given a 3-coloring of $G$, we can embed $C_G$ as a generalized terrain boundary.
\end{proof}
\subsection{Non-oriented curves}

\begin{figure}[t]
\centering
\includegraphics[width=0.6\textwidth]{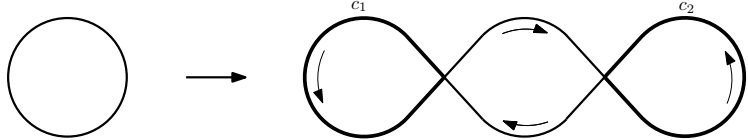}
\caption{A curve that can only be embedded with one orientation.}
\label{fi:hole}
\end{figure}

To prove a similar hardness result for unoriented curves, we simply replace each vertex-curve in the reduction by a curve that can only be embedded with one orientation, namely a curve with two self-intersections as depicted in Figure~\ref{fi:hole}. If the self-crossing parts of these curves do not cross the other curves, they can only be oriented in such a way that the two outer lobes $c_1$  and $c_2$ of the curve act like hole boundaries in whatever surface they bound, i.e. for any immersed surface $i$ the neighborhoods of $c_1$ and $c_2$ are mapped outside of $i(c_1)$ and $i(c_2)$ correspondingly.  Hence,
\begin{corollary} It is NP-complete to determine whether a curve can be seen as a boundary of an immersed surface in the plane or an embedded surface in space.
\end{corollary}
We omit the details.

\subsection{Finding a surface embedding for a single component curve}

We finish this section by showing that it is NP-complete to determine the existence of an immersed or embedded surface for a curve even when it consist of a single component. As before, we reduce the problem from {\planarcolor}. Our reduction is as before, forming a system $C_G$ of oriented curves that can be the boundary of an immersion or embedding if and only if the given graph $G$ is 3-colorable; by adding an additional step to the reduction we transform $C_G$ into a single curve without changing the existence of an immersion or embedding. The following lemmas make this extra step possible.
\begin{lemma} Let $C$ be a curve or set of oriented curves, and let $C'$ be a curve or set of curves formed by breaking $C$ in two points along the boundary of a region of thickness one, and connecting these two breaks by a pair of parallel curves. Then the embedded or immersed surfaces for C are in 1-1  correspondence with those for $C'$.
\label{le:cut}
\end{lemma}
\begin{proof}
In terms of the embedded or immersed surface, going from $C$ to $C'$ corresponds to cutting the surface by removing a thin strip between the parallel curves, and going from $C'$ to $C$ corresponds to gluing the cut back together. Cutting and gluing in this way are inverse operations, so both are one-to-one.
\end{proof}
\begin{figure}[b]
\centering
\includegraphics[width=0.5\textwidth]{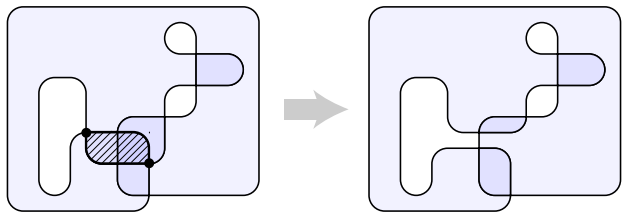}
\caption{Uncrossing the dashed region bounded by two thick segments of curves and two indicated crossing points.}
\label{fi:cut-uncross}
\end{figure}
\begin{lemma} Let $C$ be a curve or set of oriented curves, and let $R$ be a region bounded by two segments of curves and two crossing points, such that at least one of the two neighboring regions of $R$ has smaller thickness than $R$, and such that $R$ contains no crossing of $C$. Let $C'$ be the curve or set of oriented curves formed by uncrossing the two  crossings bounding $R$. Then $C$ is the boundary of an embedded or immersed surface if and only if $C'$ is.
\label{le:uncross}
\end{lemma}
\begin{proof}
In terms of the embedded or immersed surface, the two curves bounding $R$ must be boundaries of two different layers of the surface, so one can go from $C$ to $C'$ by pulling these layers apart without changing the existence of an immersion or embedding (Figure~\ref{fi:cut-uncross}). In the other direction, we can push the two layers together without any interaction between them; note however that for embeddings these two operations may not be one-to-one as we may have a choice whether to put one layer above or below the other.
\end{proof}
Given an instance $G$ of \planarcolor we construct the curve family $C_G$ as in Section~\ref{se:oriented}. We then transform $C_G$ into a single curve by attaching the three outer curves together, and the holes to the outer curve, via thin strips that pass across the curve without covering any crossings, go outside the outer boundary, and then connect back to it, as shown in Figure~\ref{fi:embedcurve-NP}(c). Each strip connecting two boundary curves can be removed by cutting it and performing a sequence of pulling-apart operations, so by Lemmas~\ref{le:cut}~and~\ref{le:uncross} the single curve is the boundary of an immersed or embedded surface if and only if $C_G$ was. Hence we have the following theorem:
\begin{theorem} The problem of determining the existence of an immersed or embedded surface that has a single component curve as a boundary is NP-complete.
\end{theorem}

\section{Finding an embedding from a surface immersion}

As we now show, it is NP-complete to lift an immersion to an embedding. Our reduction is via a graph-theoretic problem, {\acyclicpart}, which we also show to be NP-complete.
Define {\acyclicpart} to be the decision problem that takes a directed graph $G$ as input and outputs yes if and only if the vertices of $G$ can be partitioned into two acyclic subsets. We first reduce {\acyclicpart} to our  problem by constructing an immersed disk that can be lifted into space if and only the input graph is a yes-instance of \acyclicpart. Then we show that \acyclicpart is NP-complete.
\begin{figure}[t]
\centering
\includegraphics[width=0.8\textwidth]{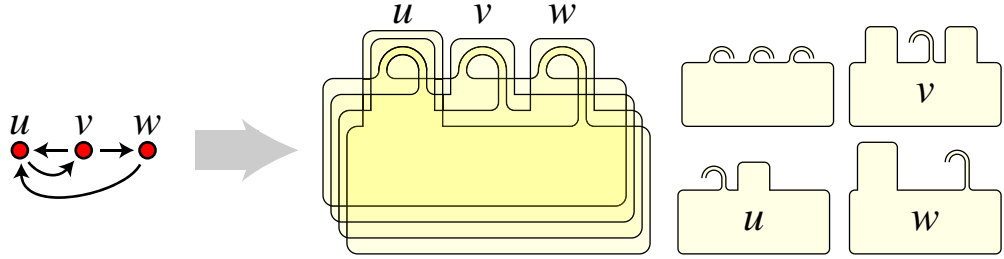}
\caption{Reduction from \acyclicpart to liftability of an immersed surface.}
\label{fi:graph2curves}
\end{figure}
\begin{lemma}
\acyclicpart can be reduced in polynomial time to disk immersion lifting.
\end{lemma}
\begin{proof}
Given a directed graph $G$, we create an immersed disk, in the form of a single central area (a large rectangle) connected to a rectangle for each vertex of $G$ that covers approximately the same region of the plane as the central area (perturbed slightly so the boundaries do not overlap). For each $v$, the rectangle for $v$ is connected to the central rectangle by a semicircular bridge that extends out from the central rectangle and back in to the rectangle for $v$, as shown in Figure~\ref{fi:graph2curves}; no two of these bridges overlap. Additionally, whenever $G$ has an edge $v \rightarrow w$, we extend a rectangular tab out from the rectangle for $v$ so that it covers the bridge for $w$. See Figure~\ref{fi:graph2curves} showing the complete reduction for a very simple graph.

How can the resulting immersed disk be embedded? Each vertex's rectangle must go either above or below the central rectangle, clearly, and as the rectangles all have a common intersection they can be totally ordered by height in any embedding. If $v$ and $w$ are both above the central rectangle, and there is an edge $v \rightarrow w$, then the corresponding tab forces $v$ to be above $w$. Thus, the total order of the rectangles above the central rectangle is consistent with the edges. This is only possible if the edges connecting pairs of points corresponding to the rectangles above the central rectangle form a directed acyclic graph. Similarly the rectangles below the central rectangle can be embedded in such a way that their tabs do not block their bridges only if the corresponding vertices form a directed acyclic graph.

In the example in Figure~\ref{fi:graph2curves}, for instance, we can embed the rectangles in top-to-bottom order $w$, $u$, central, $v$. The two rectangles above the central rectangle, $w$ and $u$, define an acyclic graph as does the rectangle $v$ by itself below the rectangle. However it would not work to try to put the rectangles for $u$ and $v$ on the same side of the central rectangle as each other, as the two of them form a cycle in the graph.
\end{proof}
Next, as promised, we prove that \acyclicpart is NP-complete.
\begin{figure}[b]
\centering
\includegraphics[width=0.3\textwidth]{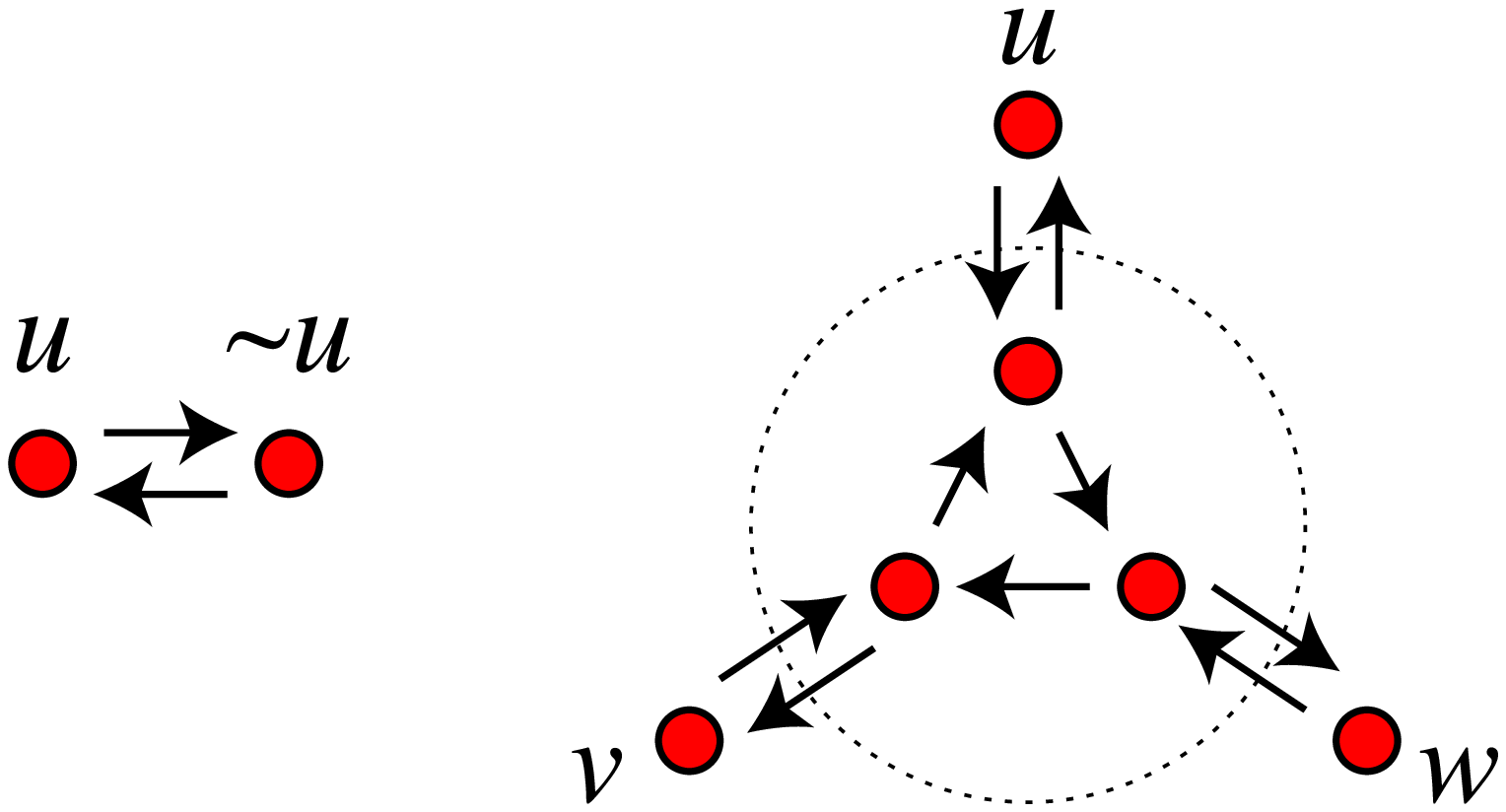}
\caption{Gadgets for the reduction from \naesat to \acyclicpart.}
\label{fig:3sat-gadgets}
\end{figure}
\begin{lemma}
\acyclicpart is NP-complete.
\end{lemma}
\begin{proof}
\naesat is a known-NP-complete variation of \sat. A \naesat instance consists of a set of clauses, each having three terms (variables or negations of variables). It is satisfied by a truth assignment such that in every clause not all three terms have the same values.

We transform \naesat into \acyclicpart as follows. Create a pair of graph vertices for each variable and its negation, with edges forming a 2-cycle (Figure~\ref{fig:3sat-gadgets}, left); the two vertices must be in different parts of any acyclic bipartition. Create another triple of vertices for each clause, with edges forming a 3-cycle; not all three may be on the same side of any acyclic bipartition.  Finally, add 2-cycles connecting the term vertices with corresponding clause vertices (Figure~\ref{fig:3sat-gadgets}, right). The resulting graph has an acyclic bipartition if and only if the input \naesat instance is satisfiable.
\end{proof}

Combining these two reductions gives us our result:

\begin{theorem}
It is NP-complete to determine whether an immersed surface can lift to a generalized terrain.
\end{theorem}

\section{The number of embeddings of an immersion}
\begin{figure}
\begin{minipage}[t]{.25\textwidth}
\centering\includegraphics[width=0.8\textwidth]{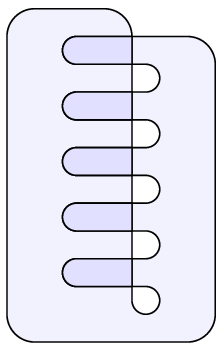}
\caption{An immersed surface with $2^{n/2}$ combinatorially different spatial embeddings.}
\label{fig:comb}
\end{minipage}
\hfill
\begin{minipage}[t]{.64\textwidth}
\centering\includegraphics[width=0.64\textwidth]{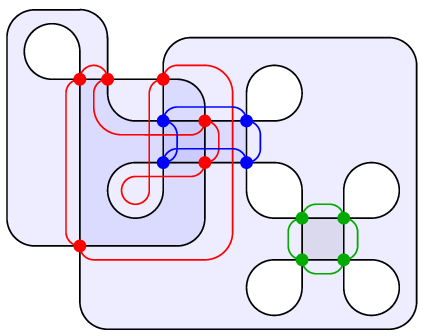}
\caption{The graph $G$ formed from the crossings of an immersed surface. The curve shown in black is the boundary of a unique immersed surface; the corresponding graph consists of two 4-cycles and one 6-cycle (colored red, green, and blue).}
\label{fig:crossing-graph}
\end{minipage}
\end{figure}

Figure~\ref{fig:comb} shows an immersed surface with $2^{n/2}$ combinatorially different spatial embeddings. As we now show, this is the most possible.
\begin{theorem}
An immersed surface $M$ has at most $2^{n/2}$ combinatorially different spatial embeddings.
\end{theorem}
\begin{proof}
We construct an undirected multigraph $G$ on a set of self-intersections of the boundary curve, as follows. From each self-intersection $v$, formed by the crossing of the boundaries of two layers $\ell_i$ and $\ell_j$, trace a curve in $\ell_j$ along the path formed in the plane by the boundary of $\ell_i$, until reaching another crossing point $w$ that involves a boundary of the layer in which the curve is being traced; add an edge in $G$ connecting $v$ and $w$. Figure~\ref{fig:crossing-graph} depicts this construction for a curve that is the boundary of a unique immersed surface.

The obtained graph $G$ is 2-regular: there is one edge for each layer involved in each crossing. If $M$ is obtained from a generalized terrain, both crossings connected by any edge $e$ of $G$ must be cased the same way: for, the path traced out by $e$ has at all points of the path one surface consistently above or below the boundary curve of the other, so there is no way for the two surfaces to swap heights. Therefore, for a casing coming from an embedding of $M$, the edges around any cycle of $G$ must alternate between upper and lower, so each cycle must have even length at least equal to two. Choosing one of two casings for each of at most $n/2$ cycles in $G$, the total number of valid casings is at most $2^{n/2}$.
\end{proof}

One consequence of this result is a simple $O(n2^{n/2})$-time algorithm for finding a lifting of an immersed surface with a single boundary component, by testing all casings consistent with the graph~$G$. \red{However, in many cases we can do better than this.}

\red{Observe that, in Figure~\ref{fig:crossing-graph}, the green cycle on the left consists of vertices and boundary paths that all lie on points contained in only two levels of the immersed surface. We call such a cycle an \emph{irrelevant cycle}, because the existence of a three-dimensional embedding does not depend on how it is cased: if a three-dimensional embedding with one of the two possible casings exists, then changing the casing of that cycle while keeping the casing of the rest of the drawing fixed will result in another valid embedding. The red and blue cycles in the figure are \emph{relevant cycles}, because they are not irrelevant: they both include vertices that are contained within three layers of the immersion. It is possible for a cycle to have all its vertices contained in only two layers, but to have points on the curves forming its edges that are contained in three or more layers; in this case, also, we call it a relevant cycle. Then clearly, we can find a lifting of an immersed surface in time $O(n2^k)$, where $k$ is the number of relevant cycles: try all casings of the relevant cycles, and pick a single fixed casing for each irrelevant one. Thus, the problem of lifting an immersion to an embedding is fixed-parameter tractable, with $k$ as the fixed parameter.}

\red{Alternatively, we may use as a parameter the number $c$ of crossings of the input curve that lie within three or more layers of the immersion. It is not difficult to show that $k=O(c)$; for, each relevant cycle either contains such a crossing or contains a two-layer crossing that is adjacent to a three-layer crossing in the arrangement formed by the boundary curves.}

\red{It would be of interest to determine whether there is a fixed parameter tractable algorithm for the problem of finding a three-dimensional embedding for a given plane curve, without being given the corresponding immersion. In this case the multigraph $G$ cannot be defined, since it depends on having a fixed immersion, but we may still use $c$ as the parameter for testing fixed parameter tractability.}

%---------------------------- Bibliography ------------------

\vfill\eject
\bibliographystyle{abbrv}
\bibliography{curves}
%\mycom{references}
\end{document}